\documentclass[a4,showpacs,preprintnumbers,amsmath,amssymb,preprint]{revtex4}

\usepackage{graphicx}%

\begin{document}
\title{Multimodal transition and excitability of a neural oscillator}

\author{L.~S.~Borkowski}
\affiliation{Department of Physics, Adam Mickiewicz University,
Umultowska 85, 61-614 Poznan, Poland}

\begin{abstract}
We analyze the response of the Morris-Lecar model to a periodic
train of short current pulses in the period-amplitude plane.
For a wide parameter range encompassing both class 2 and class 3 behavior in the Hodgkin's
classification there is a multimodal transition between the set of odd modes
and the set of all modes. It is located between the 2:1 and 3:1
locked-in regions. It is the same dynamic instability as the one discovered earlier
in the Hodgkin-Huxley model and observed experimentally in squid
giant axons. It appears simultaneously with
the bistability of the states 2:1 and 3:1 in the perithreshold regime.
These results imply that
the multimodal transition may be a universal property of resonant neurons.
\end{abstract}
\maketitle

\section{Introduction}
In 1948 Hodgkin studied the response of neurons to stimulation
by a constant current \cite{Hodgkin1948}. He summarized his findings by dividing
neurons into three classes: those having continuous relation
between the current amplitude and response frequency (type 1),
those with a discontinuous jump of response frequency at the stimulus
threshold (type 2), and those which spike only once or twice
to the constant current stimulus (type 3).
Most mammalian neurons are believed to be of type 1 \cite{Wilson1999}.
Some models in this category are
the Connor model of molluscan neurons \cite{Connor1977,Ermentrout1996},
the theta neuron \cite{Ermentrout1986,Hoppensteadt1997,Gutkin1998},
the Wang-Buzsaki model of hippocampal interneurons \cite{Wang1996}, and the Wilson-Cowan model
of a relaxation oscillator \cite{Hoppensteadt1997}.
Well known examples of type 2 neurons include Hodgkin-Huxley (HH) \cite{HH1952},
fast-spiking cortical cells \cite{Tateno2004,Tateno2006},
Morris-Lecar (ML) \cite{Morris1981,Rinzel1998}
and Hindmarsh-Rose \cite{Hindmarsh1984} models.
Some neuron models are known to exhibit different types of excitability,
depending on parameter values.
Prescott et al. \cite{Prescott2008} showed in the ML model,
originally used to describe the barnacle giant muscle fiber,
that change of one parameter was sufficient
to switch between type 1, type 2, and type 3 dynamics.

There is more evidence that the spike initiation mechanism
is not a fixed property of the neuron. In some experiments
the squid giant axons had type 3 instead
of type 2 excitability \cite{Clay1998,Clay2005}.
The discrepancy between the type 2 behavior of the HH model
and experiment was explained by modifying a single
parameter in the term describing the potassium current \cite{Clay2008}.
In a recent study
of the periodically stimulated HH model by the present author
it was found that the firing
rate may be either continuous or discontinuous function
of the current amplitude, depending on the stimulus frequency \cite{Borkowski2011}.
Bistable behavior at the excitation
threshold appears at non-resonant frequencies \cite{Borkowski2011}.
When brief stimuli arrive at resonant frequencies,
the HH neuron may respond with arbitrarily low firing rate.
The dependence of the firing rate on the current amplitude
scales with a square root above the threshold,
consistent with a saddle-node bifurcation \cite{Strogatz1994}.

The HH neuron's response at resonant frequencies
can be divided into three regimes: (i) short pulses, where the width
$\tau$ does not exceed the optimal width $\tau_{opt}$ associated with a minimum
threshold $\tau_{opt}$, (ii) $\tau_{opt} < \tau < T_{res}$,
and (iii) $\tau \simeq T_{res}$ \cite{Borkowski2011}. $T_{res}$
is defined here as the stimulation period for which the amplitude of the membrane
potential oscillations is maximum. The inverse of $T_{res}$ is the neuron's natural
frequency.
The Hodgkin classification scheme is related to case (iii).
However the analysis of response in the limit of short stimuli
gives an alternative information about
the neuron's dynamics \cite{Borkowski2011},
where a multimodal transition (MMT), involving the change of parity
of response modes, was discovered
at frequencies above the main resonance frequency \cite{Borkowski2009}.
Experimental data of Takahashi et al. \cite{Takahashi1990}
provide strong evidence for the existence of this transition \cite{Borkowski2010}.
The MMT occurs just above the threshold,
between the locked-in states 2:1 and 3:1.
Could this be a universal property of resonant neurons?
How does the MMT relate to the Hodgkin's classification?
Is it possible to use the MMT as a basis for distinguishing
between different types of neurons?
Answering these questions should increase our understanding
of the role played by various groups of neurons
in encoding different types of neural
input \cite{Tateno2004,Tateno2006,StHilaire2004,Naud2008}.
In the following we try to establish the link between
the global bifurcation diagram
of the ML model and the MMT for a parameter set used in Ref. \cite{Prescott2008}
and analyze the evolution of excitability patterns
as a function a single parameter.

\section{The model and results}

We use the form of the ML model proposed by Prescott et al. \cite{Prescott2008},

\begin{eqnarray}
\label{ML}
C\frac{dV}{dt}&=&-g_{fast}m_{\infty}(V)(V-E_{Na})\\
&-&g_{slow}w(V-E_K)-g_L(V-E_L)+I_{app},
\end{eqnarray}

\begin{equation}
dw/dt=\phi_w \frac{w_\infty(V)-w}{\tau_w(V)},
\end{equation}

\begin{equation}
m_{\infty}(V)=0.5\left[1+\tanh\left(\frac{V-\beta_m}{\gamma_m}\right)\right],
\end{equation}

\begin{equation}
w_{\infty}(V)=0.5\left[1+\tanh\left(\frac{V-\beta_w}{\gamma_w}\right)\right],
\end{equation}

\begin{equation}
\tau_w(V)=1/\cosh\left(\frac{V-\beta_w}{2\gamma_w}\right).
\end{equation}
The fast activation variable $V$ competes with the slow recovery variable $w$.
Parameter values were chosen in Ref. \cite{Prescott2008} to produce different
spiking patterns:
$E_{Na}=50 \textrm{mV}$, $E_K=-100 \textrm{mV}$, $E_L=-70 \textrm{mV}$,
$g_{fast}=20 \textrm{mS/cm}^2$, $g_{slow}=20 \textrm{mS/cm}^2$,
$g_L=2\textrm{mS/cm}^2$, $\phi=0.15$, $\beta_m=-1.2 \textrm{mV}$, $\gamma_m=18 \textrm{mV}$,
and $\gamma_w = 10 \textrm{mV}$.
$C=2\mu\textrm{F/cm}^2$ is the membrane capacitance.
We chose the input current to be a periodic set
of rectangular steps of period $T_i$, height $I_0$ and width $\tau=0.5 \textrm{ms}$.
Studying the HH model, we learned that the topology
of the global bifurcation diagram is only weakly dependent
on shape details of individual pulses,
provided they remain short compared to the time scale
of the main resonance \cite{Borkowski2011,Borkowski2011a}.
The calculations are carried out within the fourth-order Runge-Kutta scheme
with the time step of $0.001\textrm{ms}$. Individual runs at fixed parameters
were carried out for $1000 T_i$.
Since the variation of $\beta_w$ is sufficient to alter the excitability
type of the model, we study the effect of $\beta_w$ on
the dynamics at finite frequencies.
Changes of other parameters, $\beta_m$, $g_{fast}$, $g_{slow}$, $\gamma_m$,
and $\gamma_w$, may result in similar evolution
of the neuron's dynamics \cite{Prescott2008}.

\begin{figure}[th]
\includegraphics[width=0.55\textwidth]{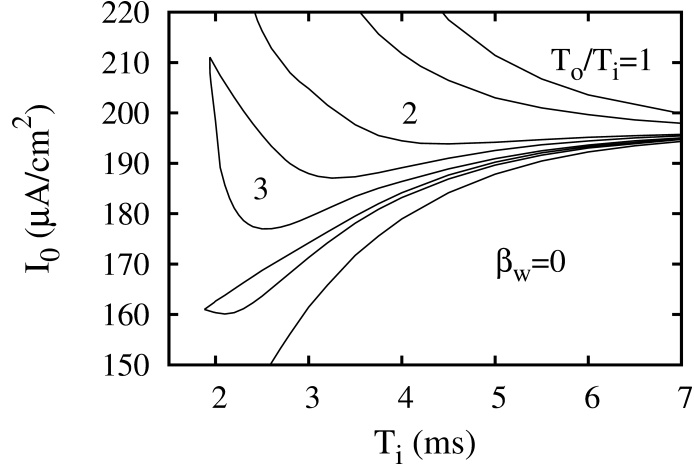}
\caption{Response diagram for $\beta_w=0$. The main locked-in
states, 1:1, 2:1, and 3:1, are labeled by the inverse of the firing rate,
$T_o/T_i$, where the $T_o$ is the average time between voltage spikes. The 4:1
state is also shown but its label is omitted due to a lack of space.
The lowest line is the excitation threshold between quiescence and a finite firing rate.
}
\label{fig1}
\end{figure}

Figure \ref{fig1} shows the global bifurcation diagram in the period-amplitude plane
for $\beta_w=0$. We call it a response diagram since it characterizes the response of
the dynamical system to a periodic perturbation. The lines on this graph
are borders between the dominant locked-in states and regions of irregular response.
In the limit, where $I_{app}$ is always constant and $T_i=\tau$, this choice
leads to type 1 excitability \cite{Prescott2008}.
In Fig. \ref{fig1} the dependence of the firing rate $f_o/f_i$ on $I_0$, where $f_i=1/T_i$,
and $f_o=1/T_o$, is continuous everywhere along the excitation threshold,
scaling approximately as $(I_0-I_{th})^{1/2}$, where $I_{th}$ is the value
of $I_0$ at the threshold. 

\begin{figure}[ht]
\includegraphics[width=0.55\textwidth]{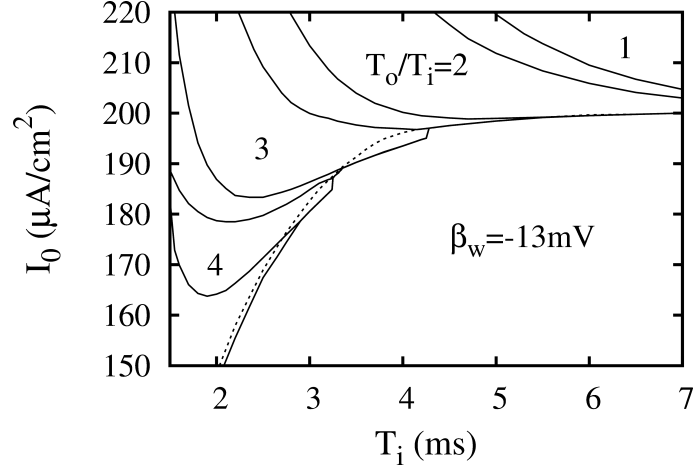}
\caption{Response diagram for $\beta_w=-13\textrm{mV}$.
The locked-in states of order 3:1 and higher are bistable along
the threshold. However the 2:1 state remains monostable.
The dashed line separates the monostable regime from the bistable area.
We have verified that MMT does not appear in this case.
}
\label{fig2}
\end{figure}

When $\beta_w=-13\textrm{mV}$, the neuron displays class 2
dynamics for a constant current. In Fig. \ref{fig2} we can see
that the firing rate is a discontinuous function of
the current amplitude also at short stimulation periods. 
For $T_i < 4\textrm{ms}$
almost the entire threshold is bistable. However,
for $T_i > 4\textrm{ms}$ the system remains monostable.
More precisely, the bistability does not extend beyond
the 3:1 state and the edge of the 2:1 state is monostable.
The long-period response in Figs. \ref{fig1} and \ref{fig2}
is very similar. Wang et al. \cite{Wang2011} also noted
the similarity of response in this regime for a sinusoidal input.

\begin{figure}[th]
\includegraphics[width=0.55\textwidth]{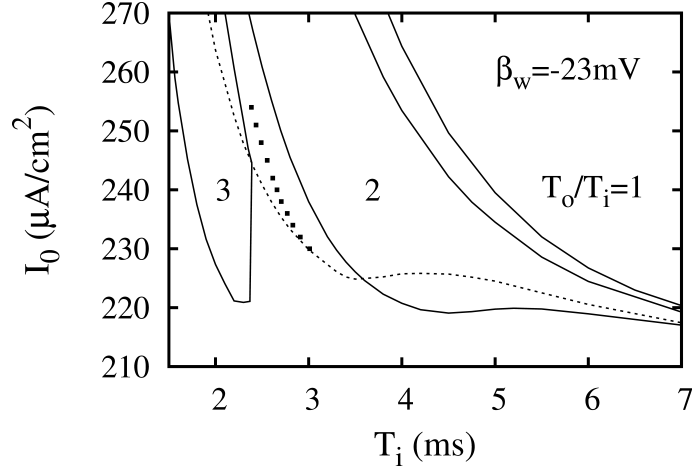}
\caption{Response diagram for $\beta_w=-23\textrm{mV}$.
The dashed line separates the monostable regime from the bistable
area. The location of the odd-all multimodal transition
is denoted by full squares. For $1.5\textrm{ms} < T_i < 5\textrm{ms}$
this diagram closely resembles the one obtained
for the HH model \cite{Borkowski2011}.
}
\label{fig3}
\end{figure}

\begin{figure}[th]
\includegraphics[width=0.55\textwidth]{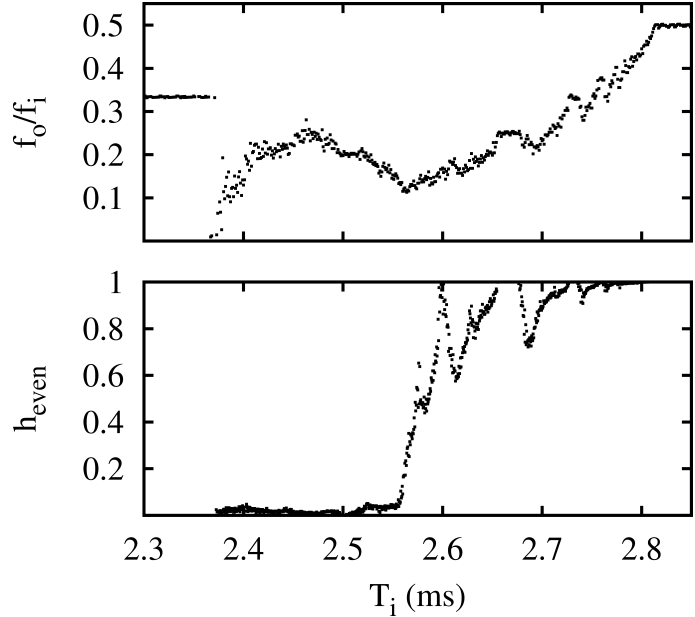}
\caption{The firing rate (top) and the weight of even modes (bottom)
for $\beta_w=-23\textrm{mV}$ and $I_0=245\mu\textrm{A/cm}^2$.
}
\label{fig4}
\end{figure}

Fig. \ref{fig3} shows the response diagram for $\beta_w=-23 \textrm{mV}$.
There are now large bistable regions of the states 3:1 and 2:1.
The short stimulation period part of the diagram, for $T_i<5\textrm{ms}$,
closely resembles the regime of the HH model where the MMT occurs
\cite{Borkowski2009,Borkowski2010}. 
We have analyzed the histogram of interspike intervals (ISI)
and found the same dynamic singularity in the ML model.
The location of the transition is indicated in Fig. \ref{fig3}
by full squares. The $f_o$ vs $I_0$ dependence in the interval
between the MMT and the 2:1 state is approximately linear, as in the HH model
\cite{Borkowski2009}.
The MMT along the $T_i$ axis for $\beta_w=-23\textrm{mV}$ is shown
in Fig. \ref{fig4}.
\begin{figure}[ht]
\includegraphics[width=0.8\textwidth]{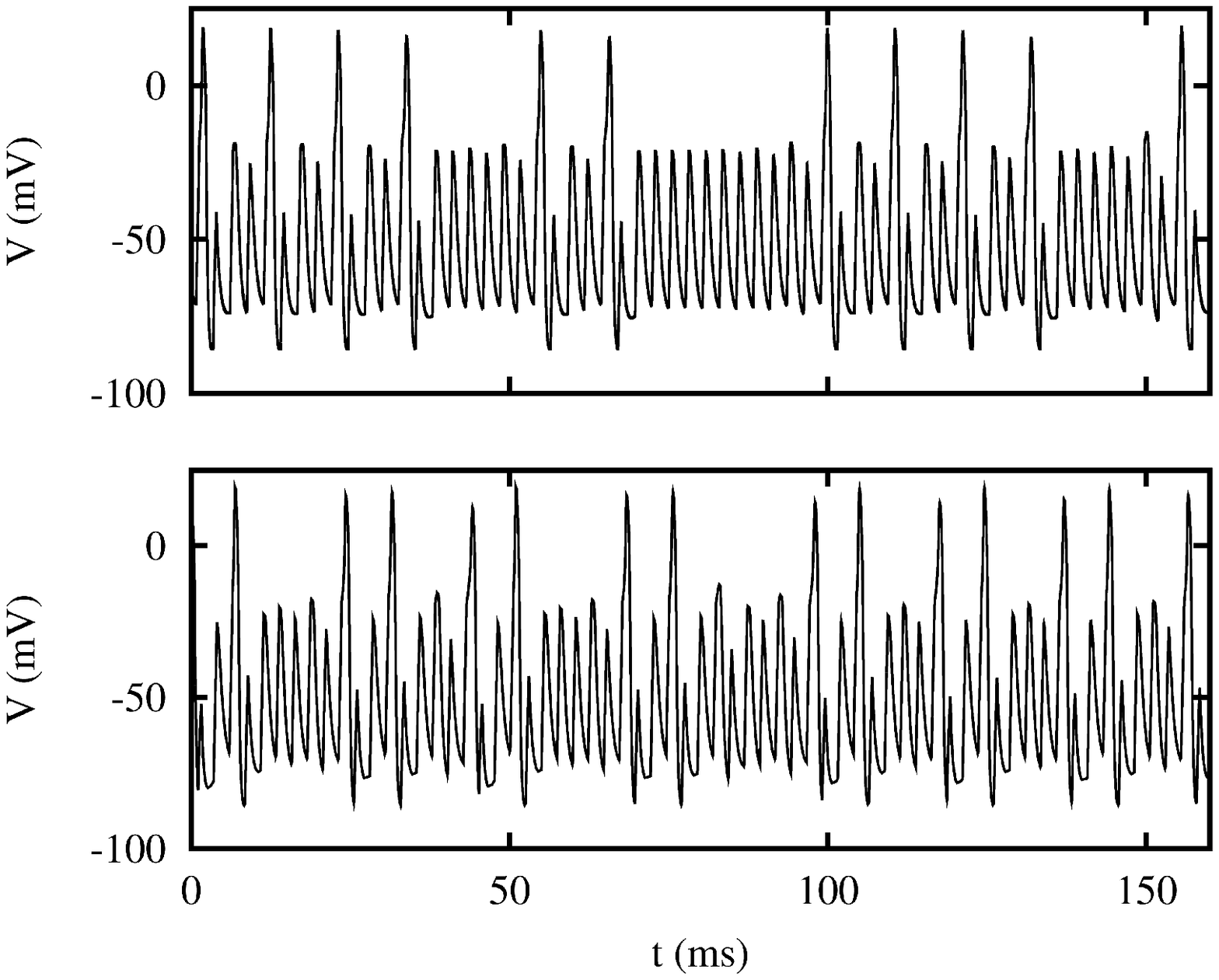}
\caption{Sample $V(t)$ run for $\beta_w=-23\textrm{mV}$, $I_0=245\mu\textrm{A/cm}^2$.
The top panel contains data taken above the odd-all transition at
$T=2.65 \textrm{ms}$. Here the even modes are more frequent
than the odd ones but there is also some admixture of the odd modes.
The bottom panel shows data at $T_i=2.45 \textrm{ms}$.
Note the absence of even multiples of the driving period in the bottom panel.
Here the response contains only the following multiples of $T_i$: 3, 5, 7, and 9. 
}
\label{fig5}
\end{figure}
The weight of even modes drops sharply
in the vicinity of the minimum of the firing rate.
The edges of individual modes near the transition
scale logarithmically, as in the HH model \cite{Borkowski2009}.
All these signatures of the MMT and the topology of the bifurcation
diagram in the vicinity of the MMT in the ML model are identical
to the HH model.

Fig. \ref{fig5} shows sample $V(t)$ runs on both sides of the even-all
transition. In the top panel, obtained at $T_i=2.65\textrm{ms}$,
the even modes 4:1 and 8:1  dominate.
In the bottom panel obtained at $T_i=2.45\textrm{ms}$
only odd modes are present.
Here the height of the $V(t)$ peak
correlates with the length of the preceding interspike interval. A careful examination
of the peak heights reveals a preference for a $T_o=3T_i$,
following the action potential with the maximum value of $V$.
This preference is manifested either as (i) another action potential,
or as (ii) subthreshold peak with a height somewhat larger
than that of its immediate neighbors. Judging by the height of subthreshold
peaks there is a clear preference for a period $2T_i$ subthreshold
oscillation. We can view the odd-only periods in the bottom panel
as a sum of $3T_i+2nT_i$, where $n=0,1,\ldots$.

\begin{figure}[ht]
\includegraphics[width=0.55\textwidth]{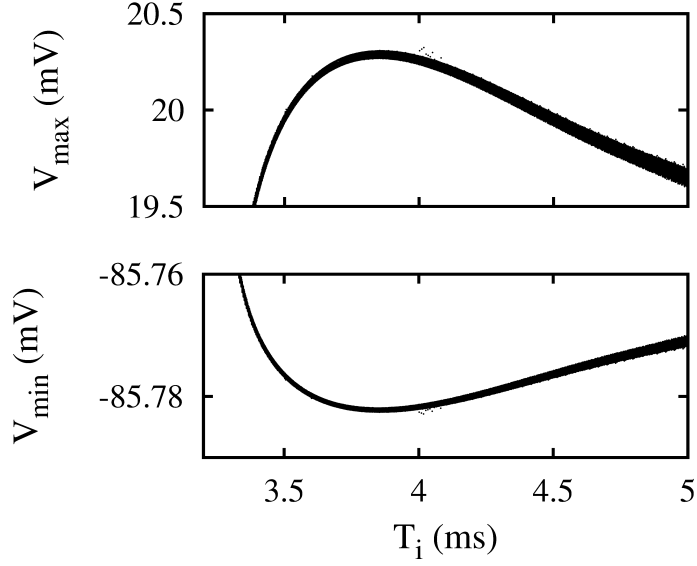}
\caption{Maximum and minimum values of $V(t)$ as a function
of $T_i$ for $\beta_w=-23\textrm{mV}$ and $I_0=230\mu\textrm{A/cm}^2$.
It is a cross section of the 2:1 mode-locked area. The optimal response
is obtained at $T_i\simeq 3.85\textrm{ms}$, which gives the resonant 
time scale of $7.7\textrm{ms}$.
}
\label{fig6}
\end{figure}

The system's resonant frequency may be estimated
by examining the extrema of $V(t)$. Fig. \ref{fig6}
shows the dependence of $V_{max}(t)$ and $V_{min}(t)$
on stimulus period in the 2:1 zone
of Fig. \ref{fig3}. The action potential amplitude reaches
maximum values at $T_i\simeq .85\textrm{ms}$. Therefore the resonant
period is $T_{res}\simeq 2\times 3.85\textrm{ms}=7.7\textrm{ms}$, giving
the resonant frequency of $f_{res}=1/7.7\textrm{ms}=130\textrm{Hz}$.
A similar estimate, although somewhat less accurate, is obtained from the minimum
of the monostable region of the 2:1 zone in Fig. \ref{fig3}
located near $T_i=3.5\textrm{ms}$.
For other values of $\beta_w$, $f_{res}$ can be obtained also from
the minimum thresholds of the 3:1 and 4:1 zones
in Figs. \ref{fig1} and \ref{fig2}.
We found that $f_{res}$ does not depend significantly on $\beta_w$.

\begin{figure}[th]
\includegraphics[width=0.55\textwidth]{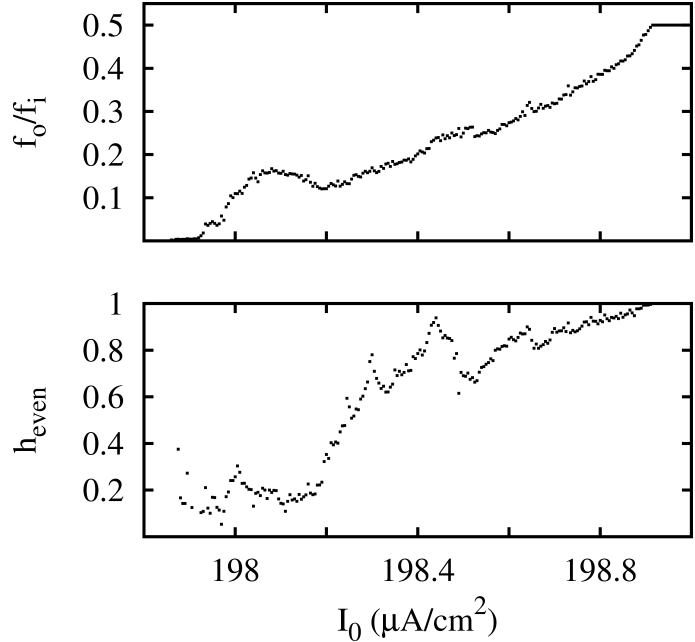}
\caption{The firing rate and the joint histogram weight
of even modes for $\beta_w=-13\textrm{mV}$ and $T_i=4.7 \textrm{ms}$.
A precursor of the multimodal transition can be seen
near $I_0=198.2 \mu\textrm{A/cm}^2$. There is a well pronounced local minimum
of the $f_o/f_i$ and a significant reduction of the participation
rate of even modes.
}
\label{fig7}
\end{figure}

We also analyzed the ISI histogram
as a function of $\beta_w$, looking for signs of the MMT. 
For $\beta_w=-13\textrm{mV}$ we can see a precursor
of the odd-all MMT in Fig. \ref{fig7}.
There is a local minimum of $f_o/f_i$ at
$I_0=198.2 \mu\textrm{A/cm}^2$ and a significant
decrease of the participation rate of even modes
close to the threshold.

\begin{figure}[th]
\includegraphics[width=0.55\textwidth]{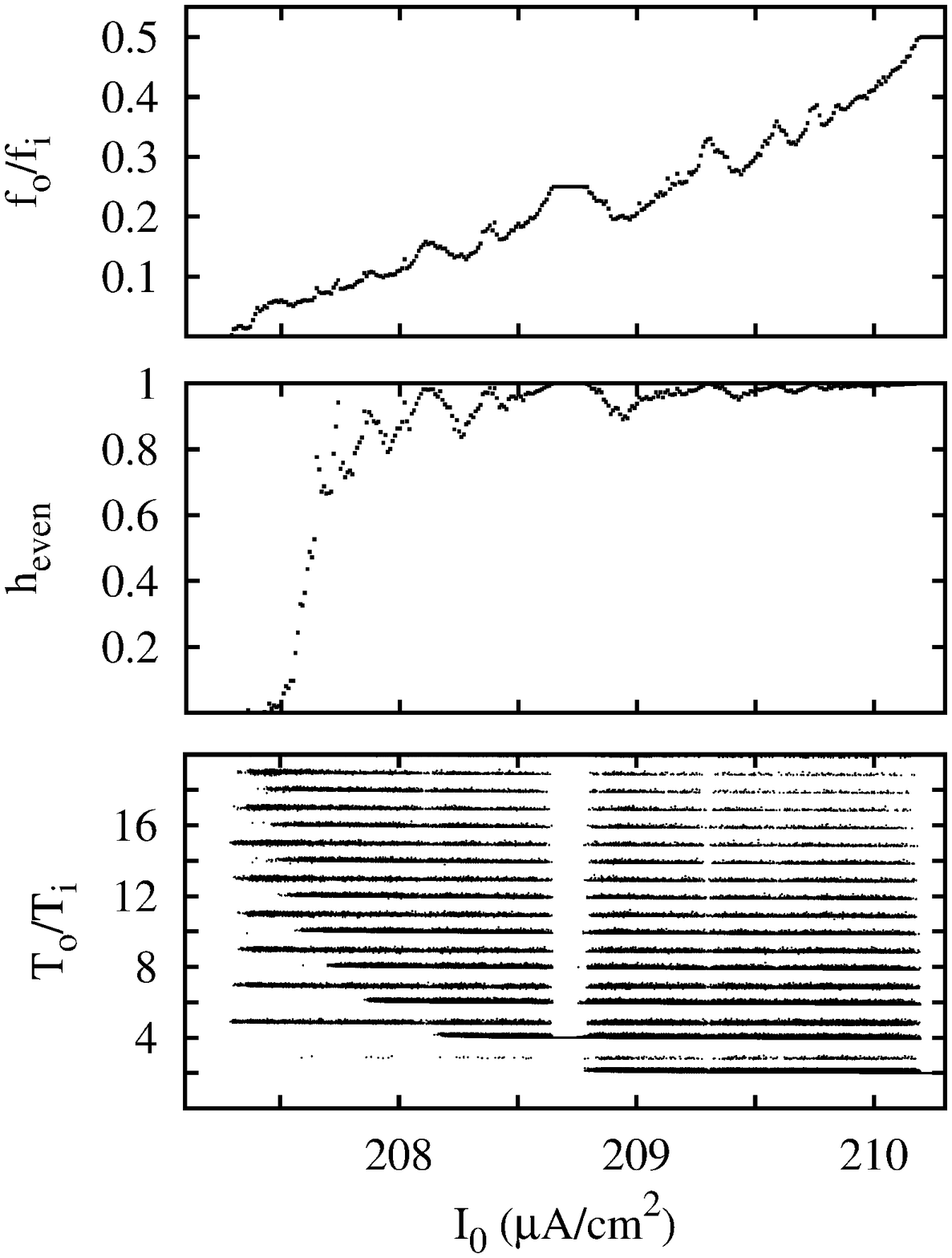}
\caption{From top to bottom: the firing rate, the joint histogram weight
of even modes, and the ISI spectrum for $\beta_w=-18 \textrm{mV}$ and $T_i=3.7 \textrm{ms}$.
The multimodal transition occurs at $I_0=207.5 \mu\textrm{A/cm}^2$.
}
\label{fig8}
\end{figure}

The MMT is tied to the appearance of bistability
along the bottom edge of the 2:1 state, which occurs
for $\beta_w < -14.5 \textrm{mV}$.
In Fig. \ref{fig8} we can see that for $\beta_w=-18\textrm{ms}$
the MMT is already well developed.
As the MMT is approached from above, only very high order
even modes remain. The edges of the even modes scale logarithmically
as a function of $|I_0-I^*|$, where $I^*$ is the current amplitude
at which the transition occurs.
When the bistability
of the state 2:1 disappears near $\beta_w=-29 \textrm{mV}$,
so does the MMT. At this value of $\beta_w$ the 3:1 state
vanishes from the global bifurcation diagram.

\section{Conclusions}

The dynamics of neurons at periods below the resonant period
is not directly related to either class 2 or class 3 excitability.
The global bifurcation diagram of the class 3 ML model
for $\beta_w=-23\textrm{mV}$ closely resembles the class 2 HH model \cite{Borkowski2009}
at intermediate periods. This is consistent with the remarks
of Prescott et al. \cite {Prescott2008}
that neurons should not be labeled as being strictly type 2 or type 3.
The transition to bistability
at the edge of a 2:1 state
in a model stimulated by a train of current pulses
does not occur for the same parameter values as the transition
to bistability for a constant current.
The bistability appears first in the limit
of small stimulation period.
As $\beta_w$ decreases further,
the perithreshold regions for larger $T_i$ also become bistable.
The MMT occurs when
both the 2:1 and the 3:1 state have bistable regions
along the threshold. The ability to predict the existence
of the MMT from the topology of the locked-in states
in the perithreshold regime is of course a very useful property.
Testing for bistabilities near the threshold is much simpler
and computationally much more efficient than analyzing
the evolution of the entire ISI histogram as a function of pulse
period or amplitude.

Wang et al. \cite{Wang2011}
noted the similarity of class 1 and class 2 neuron response
to sinusoidal signal at low frequencies.
This is not surprising, once we note, that
the threshold bistability, and a discontinuity of the firing rate $f_o/f_i$
associated with it, appears first in the limit of small $T_i$
and extends towards larger $T_i$ with decreasing $\beta_w$.
When $\beta_w < -14.5\textrm{mV}$, the 3:1 state becomes
bistable and the entire threshold
in the regime of small $T_i$ rises significantly.
The emergence of class 3 behavior may be viewed
as a strong rigidity of neuron dynamics in the limit of high stimulation period.

Based on this relationship between MMT and the global bifurcation
diagram we can classify
neuron excitability for stimuli of finite frequency:
(i) class 1, where the firing rate
is a continuous function of $I_0$ everywhere along the threshold,
(ii) class 2, with bistabilities at the edges of high-order
locked-in states, (iii) class 3, where the MMT exists, and (iv) class 4, where
both the MMT and the bistabilities are absent and the neuron responds
to a constant current by emitting only a few spikes.

The presence of the MMT in both the ML and HH model suggests
that the same transition is present in other resonant neuron
models classified as type 2 or type 3 cells.
The existence of MMT may have important physiological consequences.
It may be relevant in the high-frequency auditory nerve fiber stimulation \cite{OGorman2009}
and possibly in the clinical procedure of deep brain stimulation \cite{McIntyre2000}.



\end{document}